\documentclass[12pt]{article}
\usepackage{latexsym}

\begin{document}

\newcommand{\be}{\begin{equation}}
\newcommand{\ee}{\end{equation}}
\newcommand{\<}{\langle}
\renewcommand{\>}{\rangle}
\newcommand{\reff}[1]{(\ref{#1})}

\title{The chemical potential in the transfer matrix and in the 
path integral formulation of QCD on a lattice} 
\author{
  { Fabrizio Palumbo~\thanks{This work has been partially 
  supported by EEC under TMR contract ERB FMRX-CT96-0045}}             \\[-0.2cm]
  {\small\it INFN -- Laboratori Nazionali di Frascati}  \\[-0.2cm]
  {\small\it P.~O.~Box 13, I-00044 Frascati, ITALIA}          \\[-0.2cm]
  {\small Internet: {\tt palumbof@lnf.infn.it}}     
   }
\maketitle

\thispagestyle{empty}   % Suppress page number on front page.

%\vspace{0.2cm}

\begin{abstract}

We define the chemical potential as the Lagrange multiplier of the baryon charge operator
in the transfer matrix formalism of QCD on a lattice. Transforming the partition function
into an euclidean path integral we get the Hasenfratz-Karsh action both for Wilson 
and Kogut-Susskind fermions. In the latter case the chemical potential in the spin-diagonal
basis is half that in the flavour basis. Some open problems in the spin-diagonal basis
are pointed out.
\end{abstract}

\clearpage

\section{Introduction}

 Hasenfratz and Karsh~\cite{Hase} noticed that if in the euclidean
path integral of relativistic field theories regularized on a lattice the 
chemical potential is coupled to the Fermi fields as in classical systems, specific 
counterterms are needed which would be extremely unconvenient in numerical simulations. 
Guided by the analogy between chemical potential and gauge fields, they
exponentiated the chemical potential in the same way as the gauge fields are exponentiated 
in the link 
variables of Wilson regularization. They also showed that this definition does not require
any counterterms (in the free theory), which makes it a good candidate for numerical simulations.

From the point of view of renormalization  the way the chemical potential is introduced is 
obviously not unique~\cite{Bili}, so that one is free to look for the form which is most 
convenient for a specific purpose. But Gavai~\cite{Gava} gave an argument whereby in any case 
the chemical potential cannot couple linearly to the fermion fields in the path integral.

There are a few points which in our view should be reexamined for a fully satisfactory 
settlement
of the issue, the more so in view of 
the long standing difficulties met in numerical simulations~\cite{Hand}.

For instance, the heuristic argument does not apply with a noncompact regularization~\cite{Pal},
where the gauge fields still couple linearly to the Fermi fields. Should one in such a case 
choose a different coupling of the chemical potential?  Moreover in statistical
mechanics the chemical potential is the Lagrange multiplier of a conserved operator, 
a feature which is lost if it is exponentiated. 

A last point to be clarified concerns 
the existence of the transfer matrix for the different fermion regularizations. This is of
course a prerequisite for the lattice formulation to be correct, but its role becomes
apparent in the present approach. We start from the observation that in classical
statistical mechanics the partition function is
the trace of the exponential of (minus) the hamiltonian. Now in the corresponding formulation of 
QCD the partition function is the trace of the transfer matrix,
whose logarithm defines the hamiltonian. It is in this formulation that the chemical 
potential should behave as a Lagrange multiplier. We adopt this definition and show how
the manipulations necessary to transform the trace into a path integral change the coupling 
of the chemical potential to the Fermi fields from linear to exponential. Following this 
prescription we find in fact 
exactly the result of Hasenfratz and Karsh both for Wilson and Kogut-Susskind
fermions, with the qualification discussed below. In the latter case the value of the chemical 
potential 
in the spin-diagonal basis is half that in the flavour basis, namely their ratio is the
inverse of that of the cutoffs.
 
In our derivation we assume a given form for the transfer matrix. Now for Wilson fermions
the transfer matrix was constructed by  L\"uscher~\cite{Lusc} in terms of quark-antiquark
creation-annihilation operators, whereby the
hamiltonian and the baryonic charge find the proper definition. The proof of reflection 
positivity was given 
in the gauge $U_0 = 1\!\!1$. 
This gauge fixing is too strong, but Menotti and Pelissetto~\cite{Meno} have extended the 
proof to the  gauge $U_0 \sim 1\!\!1$, where $U_0 = 1\!\!1$ with
the exception of a single time slice. 

For the case of Kogut-Susskind we have not found in the literature a transfer matrix
with the same desirable features. In the only paper on the subject~\cite{Shar} we are
aware of, the transfer matrix is constructed in the
spin-diagonal basis, where it is found necessary to introduce auxiliary fields whose relation to
quark-antiquarks is not transparent. Another feature, possibly related to the above, might
be a difficulty in the way of the application we are interested in: the
transfer matrix is linear in the creation-annihilation operators, rather than
exponential, so that the hamiltonian contains the logarithm of these operators.

We have therefore tackled~\cite{Palu} the problem in the flavour basis, and we have found that the 
formalism developed by L\"uscher~\cite{Lusc} for Wilson fermions can easily be extended to the
case of Kogut-Susskind, provided the elementary gauge variables are attached to
 the links of the blocks rather than of the lattice. This 
  excludes the possibility of getting the action
in the spin-diagonal basis from that in the flavour basis by a simple change of variables. If 
instead the elementary gauge variables are associated to the links of the lattice, 
we can transform to the spin-diagonal basis, but we meet 
with a formal difficulty and a practical complication. In fact the proof by Menotti and Pelissetto 
cannot be adapted straightforwardly, because their gauge fixing 
has in such a case a nontrivial Faddeev-Popov determinant (as explained in detail below) 
which must be taken into account 
to establish
the reflection positivity of the transfer matrix. This is the formal difficulty. The practical 
complication is the following.
 We have constructed the transfer
matrix in the flavour basis with the "minimal" gauge coupling, the one where 
the link variables  join the quark fields along the shortest path. Now it is well known 
that a gauge coupling simple in one basis becomes very complicated in the 
other~\cite{Klub}, and in fact we find an awkward result in the spin-diagonal basis.

Both shortcomings can hopefully be overcomed, but the construction
of the transfer matrix with Kogut-Susskind fermions in the spin-diagonal basis requires
further study.

Two final observations. The first is that no claim of uniqueness is made about the
coupling of the chemical potential in the path integral formulation for a given
transfer matrix. The question of uniqueness will be discussed in separate paper. 

The second is that the procedure just outlined appears also appropriate in the study of 
a system in the canonical 
ensemble, where since one has to integrate over the (imaginary) chemical potential~\cite{Redl}, the
exact way this is introduced can be of practical importance.

 The paper is organized in the following way. We first show how the present prescription reproduces 
the Hasenfratz-Karsh result for Wilson fermions, by using L\"uscher's~\cite{Lusc} construction 
of the transfer
matrix. We then apply the same procedure to Kogut-Susskind fermions, adopting the notations
of Montvay and M\"unster~\cite{Mont}, and using the transfer matrix
especially devised for the present application~\cite{Palu}.

In this paper the pure gauge part of the transfer matrix and the partition function
will be omitted because it does not play any explicit role (but with the Kogut-Susskind
fermions in the spin-diagonal basis) and the lattice spacing is set
equal to 1.

\section{Transfer matrix and chemical potential}

We start from the definiton of the grand canonical partition function at finite baryon
density according to the ordered product
\be
Z = Tr \left\{  \exp \left(  {\mu\over T} \, \hat{Q}_B \right) 
\prod_{n_0}\hat{{\cal T}}_q(n_0) \right\},  \label{trace} 
 \ee
where the temperature $T$ is the inverse of the number $N_0$ of links in the temporal direction,
$ \hat{Q}_B $ is the baryon charge operator and $ \hat{{\cal T}}_q $ is the quark transfer matrix.
The chemical potential $\mu$ is determined by the condition that the expectation value of
the baryon number be
\be
 q_B = Z^{-1} Tr  \left\{ \hat{Q}_B\exp (  \mu N_0 \, \hat{Q}_B )\prod_{n_0}
\hat{{\cal T}}_q(n_0)
  \right\}.
\ee
The transfer matrix $ \hat{{\cal T}}_q $ can be written~\cite{Lusc,Palu} in terms of an 
auxiliary operator $\hat{T}_q(n_0) $
\be 
\hat{{\cal T}}_q(n_0) =  {\cal J} ^{-1} \hat{T}_q^{\dagger}(n_0) \,
 \hat{T}_q(n_0+s), \label{transfer}
 \ee
where ${\cal J} $ is a function of the gauge fields which will be defined later and 
$s=\pm 1$ for Wilson/Kogut-Susskind fermions respectively. This difference in sign 
does not reflect any intrinsic difference, but is only due to the different conventions adopted
by L\"uscher and Montvay-M\"unster, which we maintain for easy reference. As a consequence of
these conventions the quarks propagate backwards in time with the Wilson definition and 
forwards with the Kogut-Susskind one.
 
$\hat{T}_q $ is defined in terms of quark-antiquark creation-annihilation operators  
$\hat{x}^{\dagger}, \hat{y}^{\dagger},\hat{x},\hat{y} $ acting in a Fock space. It depends 
on the time coordinate
$n_0$ only throu the dependence on it of the gauge fields. In fact the creation and
annihilation  operators do not depend on $n_0$. They depend on the spatial
coordinates ${\bf n}$ of the sites or, in the case of Kogut-Susskind in the flavour basis,
 of the blocks, and on
Dirac, flavour and color indices, $\alpha, f,c$ sometimes comprehensively represented by $I$.
 
 In the transfer matrix formalism most often one has to do with quantities at
a given (euclidean) time  $n_0$. For this reason we adopt a summation convention
over spatial coordinates and intrinsic indices at fixed time. So for instance we
will write
\be
\overline{q}(m_0) Q(m_0,n_0) q(n_0)= \sum_{{\bf m},{\bf n},I,J} 
\overline q_{{\bf m},I}(m_0)
Q_{{\bf m},I;{\bf n},J}(m_0,n_0) q_{{\bf n},J}(n_0),
\ee
where $Q$ is the quark matrix and $q$ the quark field. Q is a function of
time dependent link operators $U_{\mu}(n_0)$ which have the standard Wilson
variables $ U_{\mu}(m_0,{\bf m}) $ as spatial matrix elements
\be
\left(U_{\mu}(m_0) \right)_{{\bf m},{\bf n}} = \delta_{{\bf m},{\bf n}} U_{\mu}(m_0,{\bf m}).
\ee
In this notation the quark action $S_q$, the baryonic charge $ \hat{Q}_B $ and the 
auxiliary operator $\hat{T}_q(n_0)  $ can be written
\be
S_q  =  \sum_{m_0,n_0}\overline{q}(m_0) Q(m_0,n_0) q(n_0)
\ee
\be 
\hat{Q}_B = \hat{x}^{\dagger} \hat{x} - \hat{y}^{\dagger} \hat{y} \label{baryon}
\ee
\be
\hat{T}_q(n_0)  = 
\exp\left[ -\hat{x}^{\dagger} M(n_0) \, \hat{x}  
-\hat{y}^{\dagger} M(n_0) \, \hat{y} \right]
 \exp\left[ \hat{y} \, N(n_0) \,\hat{x} \right].   \label{tq}
\ee
Their form is the same for Wilson and Kogut-Susskind fermions
but the matrices $Q$, $M$ and $N$ are different in the two 
cases and will be specified later. The expression of $\hat{T}_q(n_0)$ is valid in 
the gauge $U_0 = 1\!\!1$ which is not admissible. In the construction
of the path integral formulation of QCD at finite baryon density we do not need to fix the gauge,
but to lighten the formalism we will nevertheless put $U_0= 1\!\!1$ and we will reinstate $U_0$ 
in the final result.
The reader can check that keeping $U_0$ in the intermediate steps one arrives at the same result,
provided some care is exercised: for instance when $U_0 \neq 1\!\!1$ the expression $  
\hat{x}^{\dagger} M(n_0) \, \hat{x}  + \hat{y}^{\dagger} M(n_0) \, \hat{y} $ appearing in 
Eq.~\reff{tq} changes and does not commute with $Q_B$ any longer.
We anticipate that $N$ is hermitean and also $M$ is hermitean in the gauge $U_0 = 1\!\!1$. 

Now we transform the trace into a Berezin integral. To
this end we introduce between the factors in Eq.~\reff{trace} the identity
\be
1\!\!1 = \int [dx^+ dx \, dy^+ dy] \exp(-x^+ x - y^+ y) |x \, y><x \, y|,
\ee
where the basis vectors
\be
|x \, y> = |\exp( - x \,\hat{x}^{\dagger}  - y \, \hat{y}^{\dagger}) >
\ee
are coherent states and the $x^+,x , y^+ ,y $ are Grassmann variables. They are labeled by the time
slice where the unit operator is introduced. For the other indices they are subject to the 
same convention as the creation and annihilation operators . 

We then use the following equations~\cite{Lusc}. First
\be
<x | \exp \left( \hat{x}^{\dagger} M \hat{x}\right)  |x'> = \exp \left( x^+ e^ M \,x' \right).
\ee
In particular 
\be
<x,y|\exp \left(\mu \,Q_B \right)|x',y'>
= \exp \left( e^{\mu} x^+ x' + e^{-\mu} y^+ y' \right),  
\ee
which shows how the chemical potential gets exponentiated going from the trace of the transfer
matrix to the euclidean path integral.

Second, for arbitrary matrices $M$ and $N$
\be
<x|\exp (\hat{x}^{\dagger} M  \, \hat{x}) \exp (\hat{x}^{\dagger} N \, \hat{x})|x'> = 
\exp \left( x^+ e^M e^N  x' \right). 
\ee

Finally, if $B=B(\hat{x}^{\dagger})$ and $C = C( \hat{x})$ are operators which depend
on $ \hat{x}^{\dagger}, \hat{x}$ only
\be
<x|B \, M \, C |x'>= B( x^+) <x| M |x'> C(x').
\ee
Using these equations we can evaluate the kernel of the transfer matrix
\begin{eqnarray}
& &<x(n_0),y(n_0)|\hat{{\cal T}}_q(n_0) |x(n_0+s),y(n_0+s)> = {\cal J}^{-1}
\exp\left( x^+(n_0) N(n_0) y^+(n_0) \right)
\nonumber\\
& & \,\,\,\,\,\,\,\,\,\,
 \cdot   \exp\left( x^+(n_0) \exp\left(-M(n_0)\right) 
\exp\left(-M(n_0+s)\right) \exp( \mu) \, x(n_0+s) \right)
\nonumber\\
& & \,\,\,\,\,\,\,\,\,\,  \cdot \exp \left(
 y^+(n_0) \exp\left(-M(n_0)\right) 
\exp\left(-M(n_0+s)\right) \exp(- \mu) \, y(n_0+s) \right)
\nonumber\\
 & & \,\,\,\,\,\,\,\,\,\,  \cdot \exp\left(y(n_0+s) N(n_0+s) x(n_0+s)\right).  
\end{eqnarray}

Collecting all the pieces we arrive at the euclidean path integral form of the partition function
\be
 Z = {\cal J}^{-1} \int [dx^+ d x \, d y^+ d y] 
\exp  S'   \label{Seff}
\ee
with the action
\begin{eqnarray}
S' &=&  \exp ( N_0 \, \mu)  x^+(1)  \, x(1+s)
+   \exp ( -  N_0 \, \mu)  y^+(1)  \, y(1+s)
\nonumber\\
 & &+ \sum_{n_0} \left\{( 1- \delta_{n_0,1}) x^+(n_0) \exp(-M(n_0)) \exp(-M(n_0+s)) \, x(n_0+s)
\right. 
\nonumber\\
& &\left.+( 1- \delta_{n_0,1}) y^+(n_0)\exp(-M(n_0)) \exp(-M(n_0+s))  \, y(n_0+s) \right.
\nonumber\\
&  & \left.+x^+(n_0) N(n_0) \, y^+(n_0) + y(n_0+s) N(n_0 + s) \, x(n_0+s) \right.
\nonumber\\
& & \left.- x^+(n_0) x(n_0) -y^+(n_0) y(n_0)  \right\}.   
\end{eqnarray}
The chemical potential appears only on the first site. We can arrive at a uniform distribution 
by the change of variables
\begin{eqnarray}
x(n_0) & &\rightarrow \exp(s \alpha(n_0)) x(n_0),\,\,\, 
x^+(n_0) \rightarrow \exp(-s \alpha(n_0)) x^+(n_0),
\nonumber\\
y(n_0) & & \rightarrow \exp(- s \alpha(n_0)) y(n_0),\,\,\,  
y^+(n_0) \rightarrow \exp(s\alpha(n_0)) y^+(n_0)
\end{eqnarray}
where
\be
\alpha(n_0)= \alpha(1) + (n_0 -1) \mu - { 1\over 2} (1+s) N_0 \mu,\,\,\,n_0 > 1,
\ee
with $\alpha(1)$ arbitrary and $\alpha(n_0) = \alpha(n_0+N_0)$ to respect the antiperiodicity
in time of the quark field.
 We thus get
\begin{eqnarray}
S' &=& \rightarrow \sum_{n_0} \exp( \mu) x^+(n_0) \exp(-M(n_0)) \exp(-M(n_0+s)) \, x(n_0+s) 
\nonumber\\
& & +\exp(-\mu) y^+(n_0)\exp(-M(n_0)) \exp(-M(n_0+s))  \, y(n_0+s)
\nonumber\\
&  & +x^+(n_0) N(n_0) \, y^+(n_0) + y(n_0+s) N(n_0 + s) \, x(n_0+s)
\nonumber\\
& & - x^+(n_0) x(n_0) -y^+(n_0) y(n_0).   \label{Seff}
\end{eqnarray}

\section{Wilson fermions }

We set $s=1$ in Eq.~\reff{Seff} and assume
\begin{eqnarray} 
M(n_0) & = & - \ln \left( \left( 2K \right)^{{1\over 2}} B^{-{1\over 2}}(n_0) 
\right),
\nonumber\\
N(n_0) &= &2 K B(n_0)^{-{1\over 2}} c(n_0) B(n_0)^{-{1\over 2}},
\end{eqnarray}
where K is the hopping parameter and
\begin{eqnarray}
B(n_0) &=& 1\!\!1  - K\sum_{j=1}^3 \left(   U_j(n_0) T^{(+)}_j + T^{(-)}_j U_j^+(n_0)\right)
\nonumber\\
c(n_0) &=& { 1\over 2} \sum_{j=1}^3 i\,\sigma_j \left( U_j(n_0) T^{(+)}_j 
- T^{(-)}_j U_j^+(n_0) \right).
\end{eqnarray}
We have introduced the translation operators $T_j^{(\pm)}$  with matrix elements
\be
\left( T_j^{(\pm)} \right)_{{\bf n}_1,{\bf n}_2} =  
\delta_{{\bf n}_2, {\bf n}_1 \pm e_j},  \label{transl}
\ee
where the unit vectors $e_{\mu}$ have matrix elements
\be
\left (e_{\mu}\right)_{\nu} = \delta_{\mu,\nu}.
\ee
Next we define the Dirac spinors $q$ by the transformation
\begin{eqnarray}
x &=& B^{{1\over 2}} P^{(+)}_0 q, \,\,\,y = B^{{1\over 2}} P^{(-)}_0 q
\nonumber\\
\overline{q} &=& q^{\dagger} \gamma_0 \label{transf}
\end{eqnarray}
where
\be
P_0^{(\pm)} = { 1\over 2} \left( 1 \pm \gamma_0 \right).
\ee
The jacobian of this transformation is the function ${\cal J}$ introduced in Eq.~\reff{transfer}.
The charge conjugation transformation is
\begin{eqnarray}
q & =& {\cal C} ^{-1} \, \overline{q}'
\nonumber\\
\overline{q} & = & - {\cal C}^T  \, q'
\end{eqnarray}
where, with L\"usher's conventions
\be
{\cal C}= \gamma_0 \gamma_2. 
\ee
It exchanges $x$ and $y$ according to
\begin{eqnarray}
x & =&  \gamma_2 \, y'
\nonumber\\
y & = & \gamma_2 \, x',
\end{eqnarray}
justifying our definition~\reff{baryon} of the baryon charge. Now we reinstate $U_0$.
The partition function takes the form
\be
Z= \int [d \overline{q} d q] \exp S_q,
\ee
with the action
\begin{eqnarray}
S_q &=& \sum_{n_0} \left\{ \exp (\mu)2 K \overline{q}(n_0) 
 P_0^{(+)} U_0(n_0) q(n_0+1) \right.
\nonumber\\
 & & \left. + \exp(-\mu) 2K \overline{q}(n_0+1) P_0^{(-)}
  U_0^{(+)}(n_0) q(n_0)  \right.
\nonumber\\
& & \left. +\overline{q}(n_0) \left[ 2K C(n_0)  - B(n_0)\right] q(n_0) \right\}
\end{eqnarray}
where
\be
C(n_0) ={ 1\over 2} \sum_{j=1}^3 \gamma_j \left( U_j(n_0) T^{(+)}_j - T^{(-)}_j U_j^+(n_0) \right).
\ee
This is the Hasenfratz-Karsch action for Wilson fermions with Wilson parameter $r=1$. 
Notice the "plus" sign in the exponential of the action, to comply with L\"uscher's convention.

\section{Kogut-Susskind fermions in the flavour basis}

The Kogut-Susskind fermions can be defined in the flavour as well as in the spin-diagonal basis.
If we want to be able to transform from one basis to the other, the gauge fields 
must be defined on the links of the lattice. But if we want to stay in the flavour basis, 
as we will do in this Section, we can associate the gauge fields to the links of the blocks
and forget the lattice. 

We set $s=-1$ in Eq.~\reff{Seff} and assume~\cite{Palu}
\begin{eqnarray} 
M(n_0) &=& 0
\nonumber\\
N(n_0) &= &  \sum_{j=1}^3 \left[  \gamma_5 \otimes t_5 t_j  
+ \gamma_j \left(  P^{(-)}_j U_j(n_0)T^{(+)}_j \right. \right. 
\nonumber\\
& & \left. \left. - P^{(+)}_j T^{(-)}_j U_j^+(n_0)  
\right)\right] 
 +  { m \over K} \, 1\!\!1 \otimes 1\!\!1 +  \gamma_5 \otimes t_5 t_0  ,
\end{eqnarray}

where $m$ is the quark mass parameter and  K the hopping parameter.

In the tensor product the $\gamma$-matrices  act on Dirac indices, while the $t$-matrices  
\be
t_{\mu}= \gamma_{\mu}^T
\ee
act on flavor indices. The projection operators $ P_{\mu}^{(\pm)}$ are given by
 \be
P_{\mu}^{(\pm)}  ={ 1\over 2} \left[ 1\!\!1 \otimes 1\!\!1 \pm
\gamma_{\mu} \gamma_5 \otimes  t_5t_{\mu} \right].
\ee
The Dirac spinors $q$ are obtained by the transformation
\begin{eqnarray}
x &=& 4 \sqrt{K} P_0^{(+)} q, \,\,\,y^+ = 4 \sqrt{K} P_0^{(-)} q
\nonumber\\
\overline{q} &=& q^{\dagger} \gamma_0 
\end{eqnarray}
whose jacobian is the function ${\cal J}$ introduced in Eq.~\reff{transfer}.
The charge conjugation transformation
\begin{eqnarray}
q & =& {\cal C} ^{-1} \, \overline{q}'
\nonumber\\
\overline{q} & = & - {\cal C}^T \, q'
\end{eqnarray}
where
\be
{\cal C}= \gamma_0 \gamma_2 \otimes t_0 t_2, 
\ee
exchanges $x$ and $y$ according to
\begin{eqnarray}
x & =& \gamma_2 \otimes t_0 t_2 \, y'
\nonumber\\
y & = & - \gamma_2 \otimes t_0 t_2 \, x',
\end{eqnarray}
justifying our definition~\reff{baryon} of the baryon charge. Now we reinstate $U_0$.
The partition function takes the form
\be
Z= \int [d \overline{q} d q] \exp \left(- 16 \,  S_q \right),
\ee
with the action
\begin{eqnarray}
S_q &=& K \sum_{n_0} \left\{ \exp( - \mu)   \overline{q}(n_0)\gamma_0  
P_0^{(-)} U_0(n_0)  q(n_0+1)\right.
\nonumber\\ 
 & &\left. - \exp(\mu)  \overline{q}(n_0+1) \gamma_0 P_0^{(+)}
  U_0^+(n_0) q(n_0)  +  \overline{q}(n_0) N(n_0) q(n_0) \right\}. 
\end{eqnarray}
The factor $16$ in front of the action accounts for the fact that the volume element
with Kogut-Susskind fermions is 16 times larger than in the Wilson case. The chemical
potential appears in the same way as with Wilson fermions apart from the sign. This is due 
to the fact that with the conventions adopted the quarks propagate backwards in time
with the Wilson definition and forwards with the Kogut-Susskind one.

\section{The spin-diagonal basis}

We want to explore the possibility of using the results of the previous Section 
to construct QCD at finite density in the spin-diagonal basis.
To do this the gauge fields must be defined on the links so that the block link operators 
$W_{\mu}(n_0)$ have matrix elements
\begin{eqnarray}
\left(W_{\mu}(n_0)\right)_{{\bf m},{\bf n}} &=& \delta_{{\bf m}, {\bf n}}W_{\mu}(n_0,{\bf n})
\nonumber\\
W_{\mu}(n_0,{\bf n}) &=& U_{\mu}(2n_0,2{\bf n})U_{\mu}(2n_0 +e_{\mu},2 {\bf n}+e_{\mu}).
\end{eqnarray}
The relation of the coordinates of the blocks  $n_{\mu}$ to the coordinates of 
the sites $x_{\mu}$ (not to be confused with the Grassmann
variables of the previous Sections) is
\be
x_{\mu}=2n_{\mu} +\eta_{\mu},\,\,\, \eta_{\mu}=0,1.
\ee
Then the action in the flavor basis becomes
\begin{eqnarray}
S_q &=& K \sum_{n_0} \left\{ \exp( - \mu)   \overline{q}(n_0)\gamma_0  
P_0^{(-)} W_0(n_0)  q(n_0+1)\right.
\nonumber\\ 
 & &\left. - \exp(\mu)  \overline{q}(n_0+1) \gamma_0 P_0^{(+)}
  W_0^+(n_0) q(n_0)  +  \overline{q}(n_0) N(n_0) q(n_0) \right\} \label{refgau}
\end{eqnarray}
with
\begin{eqnarray} 
N(n_0) &= &  \sum_{j=1}^3 \left[  \gamma_5 \otimes t_5 t_j  
+ \gamma_j \left(  P^{(-)}_j W_j(n_0)T^{(+)}_j \right. \right. 
\nonumber\\
& & \left. \left. - P^{(+)}_j T^{(-)}_j W^+(n_0)  
\right)\right] 
 + { m \over K} \, 1\!\!1 \otimes 1\!\!1 + \gamma_5 \otimes t_5 t_0  .
\end{eqnarray}
 But the gauge fixing $W_0 \sim 1\!\!1$ has  
a nontrivial Fadeev-Popov determinant and
we cannot say that the transfer matrix is positive definite before its effect 
is taken into proper account. We then proceed in the following
way. We  consider first the free case and then 
the interacting case to show the nature of the practical complication
if it can be shown that transfer matrix remains positive definite in the gauge $W_0 \sim 1\!\!1$.

In this Section we do not find convenient our summation convention and therefore 
we will abandon it.

\subsection{The free case}

The spin-diagonal and the flavour bases are related by
\be
q_{{\bf n};\alpha,f}(n_0)^c= { 1 \over 8} \sum_{\eta} \Gamma^{(\eta)}_{\alpha, f}
\psi(2n + \eta)^c,
\,\,\, \eta_{\mu} =0,1,  \label{transf1}
\ee
where
\be
\Gamma^{(\eta)}_{\alpha,f}= \left(\gamma_1^{\eta_1}\gamma_2^{\eta_2}\gamma_3^{\eta_3}
\gamma_0^{\eta_0}\right)_{\alpha,f}.
\ee
Performing this transformation in Eq.~\reff{refgau} with $W_{\mu} = 1\!\!1$
we find the action in the spin-diagonal basis
\be
S_{\psi} =  \sum_{x} \left\{ m \overline{\psi}(x)\psi(x) + 
K \sum_{\mu} \alpha_{\mu}(x) \left[ \overline{\psi}(x)  \psi(x+e_{\mu}) -\overline{\psi}(x+e_{\mu}) 
\psi(x)\right]  +\delta S_{\psi} \right\}, 
\ee
where 
\be
\alpha_{\mu}(x)= (-1)^{x_1+x_2+...x_{\mu-1} }
\ee
and $\delta S_{\psi}$ is the contribution of the chemical potential. It is obtained by the 
tranformation of the corresponding contribution in the flavor basis
\begin{eqnarray}
\delta S_q & = & 16 K \sum_{n_0} \left\{-(1- \exp(-\mu)) \, 
\overline{q}(n_0-1) \gamma_0 P_0^{(-)}q(n_0) \right.
\nonumber\\
 & & \left. +(1 -\exp( \mu)) \, \overline{q}(n_0) \gamma_0 P_0^{(+)}q(n_0-1) \right\} 
 \end{eqnarray}
and is
\begin{eqnarray}
\delta S_{\psi}  &=& K \sum_{x_0\,even,{\bf x}} \left\{-(1- e^{- \mu}) \alpha_0(x) 
\overline{\psi}( x - e_0)  \psi(  x) \right.
\nonumber\\
 & &
 \left.+(1- e^{\mu}) \alpha_0(x) \overline{\psi}( x)  \psi(  x - e_0) \right\}.
\end{eqnarray}
Note that the sum extends only over the sites with even temporal coordinate.
Finally the total action in the spin-diagonal basis is
\begin{eqnarray}
  S_{\psi} & = & \sum_{ x_0 \, even,{\bf x}} \alpha_0(x)
\left\{\overline{\psi}(x) \psi( x+e_0)
 -\overline{\psi}(x+e_0) \psi( x) \right.
\nonumber\\
 & &
  \left.+ e^{- \mu} \overline{\psi}(x-e_0) \psi( x) -
 e^{\mu}\overline{\psi}(x)  \psi(x-e_0) \right\} 
\nonumber\\
& & +\sum_x \left\{ 
 K      \sum_j \alpha_j(x) \left[\overline{\psi}(x)  \psi(x+e_j) -
  \overline{\psi}(x+e_j) \psi(x_0) \right]  \right.
\nonumber\\
& & \left. + m  \overline{\psi}(x_0)\psi(x_0)   \right\}.
\end{eqnarray}
The chemical potential appears only on half the sites. We get a uniform coupling by 
the following change of variables in the sites with even temporal coordinate
\be
\psi(n) \rightarrow \psi(n) \, \exp(\mu / 2), \,\,\,
\overline{\psi}(n) \rightarrow \overline{\psi}(n) \, \exp(- \mu /2),\,\,\,n_0\,\,even. \label{change}
\ee
The quark action assumes then the standard form
\begin{eqnarray}
& & S_{\psi}=\sum_{x} \left\{  K  \alpha_0(x) \left[ \exp(- \mu / 2)
\overline{\psi}(x)  \psi(x+e_0)  - \exp(\mu /2)\overline{\psi}(x+e_0) \psi(x) \right] \right.
\nonumber\\
& &  + K \sum_j   \alpha_j(x)  \left.    \left[ 
\overline{\psi}(x) \psi(x +e_j) - \overline{\psi}(x+e_j) \psi(x_0) \right] 
  + m \overline{\psi}(x) \psi(x) \right\}.  \label{standard}
\end{eqnarray}
The chemical potential now appears in all the sites in the Hasenfratz-Karsh form,
but its value is halved.

\subsection{The interacting case}

When the gauge fields are present we must change the transformation~\reff{transf1} 
to make it  gauge covariant. This can be done by introducing a string of link variables 
$V_{\eta}(n)$ connecting the site $2n$ to the site $2n +\eta$
\be
q_{n;\alpha,f}^c= { 1 \over 8} \sum_{\eta} \Gamma^{(\eta)}_{\alpha, f} \left( V_{\eta}(n)
\psi(2n+\eta)\right)^c,\,\,\, \eta_{\mu} =0,1.  \label{transf2}
\ee
 A discussion of different ways to construct $V_{\eta}(n)$ can be found
in~\cite{Klub}. For instance we can assume
\begin{eqnarray}
V_{\eta}(n) &=&
\left[ U_1(2n)\right]^{\eta_1}  \left[ U_2(2n+ \delta_{1,\eta_1}e_1) \right]^{\eta_2}
\left[ U_3(2n+ \delta_{1,\eta_1}e_1 + \delta_{1,\eta_2}e_2) \right]^{\eta_3}
\nonumber\\
& & \left[ U_0(2n+ \delta_{1,\eta_1}e_1 +\delta_{1,\eta_2}e_2
+\delta_{1,\eta_3}e_3)\right] ^{\eta_0}.
\end{eqnarray}
 
The contribution of the chemical potential to the action in the presence of the gauge fields 
can be read off from Eq.~\reff{refgau}
\begin{eqnarray}
\delta S_q & = & K \sum_{n_0}\left\{-(1-  \exp(-\mu) )\, \overline{q}(n_0-1) 
\gamma_0 W_0(n_0-1)P_0^{(-)}q(n_0) \right.
\nonumber\\
& &\left. +( 1 -\exp( \mu)) \, \overline{q}(n_0) \gamma_0 P_0^{(+)}W^+_0(n_0-1)q(n_0-1) \right\}. 
 \end{eqnarray}
After the transformation~\reff{transf2} this becomes
\begin{eqnarray}
\delta S_{\psi} & = & K \sum_{n, \eta}  \alpha_0(\eta) 
\left\{ ( 1 -\exp( \mu)) \, \delta_{\eta_0,0} \right.
\nonumber\\
 & & \left.  \cdot \overline{\psi}(2n + \eta) 
V^+_{\eta}(n) W^+_0(n- e_0) V_{\eta + e_0}(n-e_0)\psi(2n + \eta - e_0)  \right.
\nonumber\\
& & \left. - (1-  \exp(-\mu) )\, \delta_{\eta_0,1} \right.
\nonumber\\
& &\left.
 \cdot \overline{\psi}(2n + \eta) 
V^+_{\eta}(n) W_0(n) V_{\eta-e_0} (n + e_0) \psi(2n + \eta + e_0) \right\}.
 \end{eqnarray}
Therefore the total action in the spin-diagonal basis is
\begin{eqnarray}
S_{\psi} &=& \sum_{n,\eta}\left\{ K \sum_{\mu}\alpha_{\mu}(\eta)\overline{\psi}(2n + \eta) 
V^+_{\eta}(n)\left[
  A_{\mu}(n,\eta)\psi(2n+\eta + e_{\mu})  \right. \right.
\nonumber\\
& & \left.\left. - B_{\mu}(n,\eta)\psi(2n+\eta - e_{\mu}) \right]
+m \overline{\psi}(2n+\eta) \psi(2n+\eta) \right\}
\end{eqnarray}
where
\begin{eqnarray}
A_0(n,\eta) &=& \delta_{\eta_0,1}\exp(-\mu) W_0(n)V_{\eta-e_0}(n+e_0)
+\delta_{\eta_0,0} V_{\eta+e_0}(n),
\nonumber\\
B_0(n,\eta) &=&  \delta_{\eta_0,0}\exp(\mu) W_0^+(n-e_0)V_{\eta+e_0}(n-e_0)
+\delta_{\eta_0,1} V_{\eta-e_0}(n),
\nonumber\\
A_j(n,\eta) &=& \delta_{\eta_j,1}W_j(n)V_{\eta- e_j}(n+e_j)
+\delta_{\eta_j,0}V_{\eta+ e_j}(n),
\nonumber\\
B_j(n,\eta) &=& \delta_{\eta_j,0}W^+_j(n-e_j)V_{\eta +e_j}(n-e_j)
+\delta_{\eta_j,1}V_{\eta- e_j}(n).
\end{eqnarray}
Again the chemical potential appears only on half the sites and we get a uniform coupling by 
the change of variables~\reff{change} which changes $A_0,B_0 \rightarrow A_0',B_0'$ according to
\begin{eqnarray}
A_0(n,\eta)' & = &  e^{-\mu /2}\left[ \delta_{\eta_0,1} W_0(n)V_{\eta-e_0}(n+e_0)
+\delta_{\eta_0,0} V_{\eta+e_0}(n) \right]
\nonumber\\
B_0(n,\eta)' & = & e^{\mu /2} \left[ \delta_{\eta_0,0} W_0^+(n-e_0)V_{\eta+e_0}(n-e_0)
+\delta_{\eta_0,1} V_{\eta-e_0}(n) \right].
\end{eqnarray}
The chemical potential now appears in all the sites in the Hasenfratz-Karsh form,
but its value is halved.

\section{Summary}

We have seen that we can couple the chemical potential linearly to the Fermi fields,
conserving its meaning of a Lagrange multiplier in the transfer matrix formalism,
which is equivalent to the hamiltonian formalism in statistical mechanics. 
Starting from this definition we have derived the
Hasenfratz-Karsh action showing that the exponentiation of the chemical 
potential is a property
of the Grassmannian kernel of fermionic operators, and is not related to the
exponential dependence of the gauge fields.

In the case of Kogut-Susskind, the value of the chemical potential depends on the basis,
and this can be important when the chemical potential has a physical interpretation, 
which in the present scheme is only possible in the flavour basis.  

The formulation of QCD at finite baryon density in the spin-diagonal basis cannot
be considered settled. Within our approach one should prove the reflection positivity
of the transfer matrix in the gauge $W_0 \sim 1\!\!1$. Then it might also be possible
to justify the minimal coupling in the spin-diagonal basis, by appropriately changing
the gauge coupling in the flavour basis.
Alternatively one should use a transfer matrix constructed directly in the spin-diagonal
basis.

 We postpone to separate paper the discussion of the
uniqueness of the path integral for a given transfer matrix.

\subsection*{Acknowledgments}

It is a pleasure to thank dr.G.DiCarlo for many conversations about QCD at
finite baryonic density.


\clearpage 

\begin{thebibliography}{9}

\bibitem{Hase} 
P.Hasenfratz and F.Karsh, Phys.Lett.125B (1983) 308
J.Kogut, M. Matsuoka,M.Stone,H.W.Wyld,J.H.Shenker,J.Shighemitsu and D.K.Sinclair, Nucl.
Phys.B225[FS9](1983)93

\bibitem{Bili}
N.Bilic and R.V.Gavai, Z. Phys. 23C (1984) 77

\bibitem{Gava}
R.V.Gavai, Phys. Rev. D32 (1985) 519 

\bibitem{Hand}
S.Hands, Nucl.Phys.B (Proc.Suppl.) 106 (2002) 142; hep-lat/0109034

\bibitem{Pal}
F.Palumbo, Phys. Lett.B 244 (1990) 55; C. M. Becchi and F.Palumbo, Nucl. Phys. B388 (1992)
595; F. Palumbo, G. DiCarlo and R. Scimia, Nucl. Phys. Proc. Suppl. 106(2002) 823 

\bibitem{Lusc}
M.L\"uscher, Commun.math.Phys. 54 (1977) 283

\bibitem{Meno}
 P.Menotti and A.Pelissetto, Comm. Math. Phys. 113 (1987) 369


\bibitem{Shar}
H.S.Sharatchandra, H.T.Thun and P.Weisz, Nucl.Phys.B192(1981)205

\bibitem{Palu}
F.Palumbo, to be published

\bibitem{Mont}
I.Montvay and G.Munster, Quantum fields on a lattice, Cambridge University Press, 1994

\bibitem{Klub}
H.Kluberg-Stern, A.Morel, O.Napoly and B.Petersson, Nucl. Phys. B220[FS8] (1983) 447;
G.T.Bodwin and E.V.Kovacs, Phys. Rev. 38D (1988) 1206


\bibitem{Redl}
K.Redlich and L.Turko, Z.Physik C5 (1980) 201; L.Turko, Phys.Lett. 104B (1981) 153 






\end{thebibliography}
\end{document}